\documentstyle[12pt]{article}
\newcommand{\Ir}{Z\!\!\!Z}
\newcommand{\Ibb}[1]{ {\rm I\ifmmode\mkern
            -3.6mu\else\kern -.2em\fi#1}}
\newcommand{\ibb}[1]{\leavevmode\hbox{\kern.3em\vrule
     height 1.2ex depth -.3ex width .2pt\kern-.3em\rm#1}}
\newcommand{\Cx}{{\ibb C}}

\newcommand{\Rl}{{\Ibb R}}
\newcommand{\be}{\begin{eqnarray}}
\newcommand{\ee}{\end{eqnarray}}
\renewcommand{\O}{\Omega}
\newcommand{\A}{{\cal A}}

\renewcommand{\d}{\mbox{d}}

\newcommand{\pa}{\partial}

\begin{document}

\begin{tabbing}
\hspace*{12cm} \= GOET-TP 1/96  \\
               \> January 1996
\end{tabbing}
\vskip1.cm

\centerline{\huge \bf Noncommutative Geometry} 
\vskip.5cm

\centerline{\huge \bf and Integrable Models}
\vskip1cm

\begin{center}
      {\large \bf Aristophanes Dimakis} \ and \ {\large \bf Folkert   
       M\"uller-Hoissen}
       \vskip.3cm 
      Institut f\"ur Theoretische Physik  \\
      Bunsenstr. 9, D-37073 G\"ottingen, Germany
\end{center} 
\vskip1cm

\begin{abstract}
\noindent
A construction of conservation laws for $\sigma$-models  
in two dimensions is generalized in the framework of noncommutative 
geometry of commutative algebras. This is done by replacing the
ordinary calculus of differential forms with other differential
calculi and introducing an analogue of the Hodge operator on the
latter. The general method is illustrated with several examples. 
\end{abstract}


\section{Introduction}
In a recent work \cite{DMH95} we have shown that completely 
integrable discrete versions of two-dimensional $\sigma$-models 
(chiral models) are obtained via certain deformations \cite{DMHS93} of  
the ordinary calculus of differential forms on $\Rl^2$. The procedure 
is based on a generalization of the construction of conserved currents
presented for continuum $\sigma$-models in \cite{Bre79}. In the 
present work we further generalize this method in several ways.
We present rather weak conditions to be imposed on a differential 
calculus and on a generalized Hodge $\ast$-operator such that the 
classical field equation $\d \, \ast g^{-1} \, \d g = 0$ for a 
$\sigma$-model makes sense and a construction of an infinite 
sequence of conserved currents still works (section 3). In section 2 
we introduce two-dimensional `noncommutative geometries' with 
several examples to which we refer in the sequel. Section 3 presents 
our notion of a generalized $\sigma$-model, the construction of 
conserved currents for it, and a linear system of which the field 
equations are integrability conditions. This linear system is then 
further discussed in section 4 from a slightly more general point
of view, revealing a kind of `duality' between the $\sigma$-model 
field equation and the zero curvature condition. Some integrable 
equations are derived from the examples of noncommutative 
geometries in section 2. Section 5 contains our conclusions and 
further remarks.

\section{Two-dimensional noncommutative geometries}
Let $\cal A$ be a commutative algebra of functions of two variables, $t$ 
and $x$. Let $\Omega({\cal A})$ be a differential
calculus on $\cal A$ such that $\d t$ and $\d x$ constitute a left
and right $\cal A$-module basis of the space $\Omega^1({\cal A})$
of 1-forms.  Though the algebra $\cal A$ itself is commutative
(and can thus be realized as an algebra of functions on some
topological space), the differential calculus may be such that
functions and differentials do not commute. In that case we 
speak of a `noncommutative differential calculus'  and geometric
structures built on it inherit this noncommutativity. In this sense
we obtain a `noncommutative geometry' on a commutative algebra.
Our basic geometric ingredient is a $\Cx$-linear operator
$\ast \, : \,  \Omega^1({\cal A})  \rightarrow \Omega^1({\cal A})$  
with the property
\be               \label{ast1}
       \ast \, ( \omega \, f ) = f  \, ( \ast \, \omega )
\ee
for all $f \in {\cal A}$ and $\omega \in \Omega^1({\cal A})$. Then
\be                \label{ast_dt_dx}
      \ast \,  \d t  = \gamma \, \d t + \alpha \, \d x   \,  , \qquad
      \ast \,  \d x  = \beta \, \d t + \delta \, \d x   
\ee
and the operator $\ast$ is determined by the choice of $\alpha,
\beta, \gamma, \delta \in {\cal A}$. The $\ast$-operator generalizes
the Hodge operator of Riemannian geometry. 
In the following sections we shall need some additional properties 
for this operator. We require $\ast$ to be {\em symmetric} in the 
sense that
\be                \label{ast2}
        \omega \ast \omega' = \omega' \ast \omega
\ee 
for all $\omega, \omega' \in \Omega^1({\cal A})$. Depending on the
choice of differential calculus, these conditions restrict the 
possibilities for the $\ast$ operator. A simple calculation shows that 
the symmetry condition is equivalent to\footnote{It is possible
to extend a first order differential calculus to higher orders
by demanding that the product of any two differentials vanishes. Then
the following condition is trivially satisfied, but also the field 
equation which we consider in section 3. It is more natural and
convenient, of course, to constrain the space of two-forms only by
those conditions which are derived from the first order calculus 
using the general rules of differential calculus (cf \cite{DMHS93},
for example).}
\be                \label{ast3}
      \d t \, f \beta \, \d t + \d t \, f \delta \, \d x 
   - \d x \, f \gamma \, \d t - \d x \, f \alpha \, \d x = 0 
\ee
for all $f \in {\cal A}$.  We also require $\ast$ to be invertible. As a 
consequence of (\ref{ast1}) its inverse then satisfies       
$ \ast^{-1} \, ( f \, \omega ) =  (\ast^{-1} \omega ) \, f $.
Moreover, we demand that 
\be                         \label{astast}
          \d ( \ast \ast \, \omega) = 0 \quad \Leftrightarrow \quad
          \d \omega = 0   \; .
\ee
In section 3 we also need the triviality of the first cohomology group
of $\Omega({\cal A})$, i.e., closed 1-forms have to be exact. This 
condition is fulfilled for all the following examples.

\vskip.3cm
\noindent
{\em Example 1.} Let $\cal A$ be the algebra of $C^\infty$-functions on
$\Rl^2$ and $\Omega({\cal A})$ the ordinary differential calculus (where
functions commute with 1-forms). According to the Poincar{\'e} Lemma
every closed form is exact, i.e., the cohomology is trivial.
The symmetry condition (\ref{ast3}) 
becomes $\delta = -\gamma$. The $\ast$-operator is invertible iff
${\cal D} := \alpha \, \beta + \gamma^2$ is everywhere different from 
zero. We find $\ast \ast = {\cal D} \, id$.  The condition 
(\ref{astast}) is satisfied iff $\cal D$ is constant. 

\vskip.3cm
\noindent
{\em Example 2.} Let $\cal A$ be the set of all functions on the 
two-dimensional lattice $a \Ir \times b \Ir$ where $a,b$ are positive 
real constants. $x$ and $t$ are the canonical coordinate functions. 
A differential calculus on $\cal A$ is then 
determined by the commutation relations\footnote{More precisely,
these relations determine a differential calculus on the algebra of
polynomials in $x$ and $t$ which can then be extended to the algebra
of arbitrary functions. See also \cite{DMHS93}.}
\be              \label{cr_lattice} 
    \lbrack \d x , x \rbrack = a \, \d x \, , \quad 
    \lbrack \d x , t \rbrack = 0 = \lbrack \d t , x \rbrack \, ,\quad
    \lbrack \d t , t \rbrack = b \, \d t  \; .
\ee
More generally, we have
\be
     \d t \, f(x, t) = f(x, t+b) \, \d t  \, , \quad
     \d x \, f(x, t) = f(x+a, t) \, \d x
\ee
for $f \in {\cal A}$. Acting with the exterior derivative on  
(\ref{cr_lattice}) leads to
\be            \label{2-form-rels}
      \d x \, \d x = 0 = \d t \, \d t \, , \quad
      \d x \, \d t = - \d t \, \d x   \; .
\ee 
In general, however, 1-forms do not anticommute in this calculus.
For the differential of a function $f$ we get
\be
      \d f = \pa_{+x} f \, \d x + \pa_{+t} f \, \d t
\ee
where $\pa_{+t}$ and $\pa_{+x}$ are discrete partial derivatives, i.e.,
\be
 (\pa_{+x} f)(x,t) = {1 \over a} \, \lbrack f(x+a, t) - f(x,t) \rbrack
                     \, , \qquad
 (\pa_{+t} f)(x,t) = {1 \over b} \, \lbrack f(x, t+b) - f(x,t) \rbrack
      \; .
\ee
The symmetry condition (\ref{ast3})  becomes $\gamma = \delta = 0$
and the $\ast$-operator is invertible iff $\alpha \beta$ nowhere
vanishes on the lattice. Furthermore, one finds 
\be
       \ast \ast \, \lbrack \d x \, f(x,t) + \d t \, h(x,t) \rbrack
    &=& \alpha(x,t) \, \beta(x, t-b) \, \d x \, f(x-a, t-b) \nonumber \\
    & & + \alpha(x-a,t) \, \beta(x, t) \, \d t \, h(x-a, t-b)  \; .
\ee
The condition (\ref{astast}) in particular requires $\ast \ast \d t$ and
$\ast \ast \d x$ to be closed. This leads to
\be            \label{ex_2a}
  \pa_{+t} \lbrack \alpha(x,t) \, \beta(x, t-b) \rbrack = 0 \, , \qquad
  \pa_{+x} \lbrack \alpha(x-a,t) \, \beta(x, t) \rbrack = 0  \, .
\ee
Thus
\be            \label{ex_2b}
   \alpha(x,t) = { C(x) \over B(t-b) } \, \alpha(x-a, t-b) \, , \qquad
   \beta(x,t) = { B(t) \over \alpha(x-a,t) }   
\ee
where $C(x)$ and $B(t)$ are arbitrary (nowhere vanishing) functions.
Taking (\ref{ex_2a}) into account, (\ref{astast}) applied to the closed 
1-form $\d x \, t + \d t \, x$ yields
\be
   \alpha(x,t) \, \beta(x, t-b) = \alpha(x-a,t) \, \beta(x, t) \; .
\ee
Together with (\ref{ex_2b}) this requires $C$ and $B$ to be constant
and, moreover, $C = B$. We end up with
\be
      \alpha(x,t) = \alpha(x-a, t-b) \, , \qquad
      \beta(x,t) = { C \over \alpha(x-a,t) }   \; .
\ee
With these restrictions on $\alpha$ and $\beta$ we have 
\be
        \ast \ast \, \omega(x,t) = C \; \omega(x-a,t-b)
\ee
for all $\omega \in \Omega^1({\cal A})$ and (\ref{astast})
is satisfied. In the limit $a \to 0, \, b \to 0$ we obtain the ordinary
differential calculus (on $C^\infty$-functions of $x$ and $t$). The 
corresponding limit of the $\ast$-operator, however, does not exhaust 
the possibilities which we have for $a=b=0$ (cf example 1). 
On the other hand, the limit $b \to 0$, keeping $a$ constant (and
different from zero), does exhaust the possibilities which one
finds by investigating the limit calculus.
 
\vskip.3cm
\noindent
{\em Example 3.} Let $\cal A$ be the algebra of $C^\infty$-functions on
$\Rl^2$ and $\Omega({\cal A})$ the differential calculus determined by
\be
    \lbrack \d x , x \rbrack = \eta \, \d t \, , \quad 
    \lbrack \d x , t \rbrack = \lbrack \d t , x \rbrack 
      = \lbrack \d t , t \rbrack = 0 \, ,
\ee
with a constant $\eta$ (see also \cite{DMH93}). More generally, we have
\be
      \d t \, f = f \, \d t  \, ,  \quad  
      \d x \, f = f \, \d x + \eta \, f_x \, \d t    
\ee
for $f \in {\cal A}$. Here $f_x$ denotes the partial derivative with
respect to $x$. Furthermore, one finds
\be 
  \d f =  ( f_t + {\eta \over 2} \, f_{xx} )  \, \d t +  f_x  \, \d x
\ee
and $\d x \, \d x = 0 = \d t \, \d t \, , \; \d x \, \d t = - \d t 
\, \d x$. For $\eta \neq 0$ the symmetry condition (\ref{ast3}) becomes 
$\alpha = 0$ and $\delta = -\gamma$ so that
\be
      \ast \,  \d t  = \gamma \, \d t  \,  , \qquad
      \ast \,  \d x  = \beta \, \d t  - \gamma \, \d x   \; .
\ee
The $\ast$-operator is invertible iff $\gamma \neq 0$. The condition
(\ref{astast}) applied to the differentials $\d t$ and $\d x$ requires
$\gamma$ to be constant. Since every 1-form $\omega$ can be written
as $\omega = \d t \, f + \d x \, h$ with functions $f$ and $h$, a direct 
calculation now leads to
\be
       \ast \ast \omega = \gamma^2 \, \omega
\ee 
so that (\ref{astast}) is indeed satisfied. There is no restriction for 
the function $\beta$.

\vskip.3cm
\noindent
{\em Example 4.} For $ab \neq 0$, the constants $a$ and $b$ in 
(\ref{cr_lattice}) can be absorbed via a rescaling of $x$ and $t$. We 
may therefore set $a=b=1$. In terms of the `light cone coordinates'
\be 
      u :=\mu \, (t+x) \, , \qquad  v :=\nu \, (t-x)  \, ,     
\ee
where $\mu,\nu$ are constants,  (\ref{cr_lattice}) becomes
\be                    \label{cr_lattice2}
      [ \d u , u ] = \mu \, \d u \, , \quad 
      [ \d u , v ] = [ \d v , u ] = \mu \, \d v \, , \quad 
      [ \d v , v ] = {\nu^2 \over \mu} \, \d u   \; .
\ee
Performing the limit $\mu \to 0$ in such a way that $\nu^2/\mu \to 
\eta$ with a constant $\eta$, the calculus of example 3 is recovered.
Another calculus, which will be discussed in the following, is obtained 
in the limit $\nu \to 0$. After a renaming of the coordinate functions 
we get
\be  
      [ \d t , t ] = 0 \, , \quad 
      [ \d t , x ] = [ \d x , t ] = \mu \, \d t \, , \quad 
      [ \d x , x ] = \mu \, \d x   \; .
\ee
For a function $f$ this generalizes to
\be
 \d t \, f(x, t) = f(x+\mu, t) \, \d t  \, , \quad
 \d x \, f(x, t) = f(x+\mu, t) \, \d x + \mu \, \dot{f}(x+\mu,t) \, \d t 
\ee
where $\dot{f} = \partial f/\partial t$. Furthermore,
\be
  \d f = \dot{f}(x+\mu,t) \, \d t + (\partial_{+x} f)(x,t) \, \d x \; . 
\ee
The algebra $\cal A$ should now consist of functions on $\mu \Ir \times 
\Rl$ which are smooth in the variable $t$. 
Again, (\ref{2-form-rels}) holds.
The $\nu \to 0$ limit of the $\ast$-operator for the calculus of example 
2 (in the form (\ref{cr_lattice2})) only leaves us with $\alpha = \beta 
=0$ and $\delta = -\gamma$ in (\ref{ast_dt_dx}). But a closer inspection 
of the above (limit) calculus shows that an arbitrary function $\beta$ 
is permitted. The condition (\ref{astast}) requires $\gamma$ to be 
constant and $\beta$ not to depend on $x$, i.e., $\beta = \beta(t)$. 
Then
\be
       \ast \ast \, \omega(x,t) = \gamma^2 \; \omega(x - 2 \mu, t)  \; .
\ee

\vskip.3cm

The above examples by far do not exhaust the 
possibilities.\footnote{Further examples of two-dimensional differential 
calculi can be found in \cite{BDMH95}.} Even these examples can be
considerably generalized by replacing the constants appearing in the
defining relations of the differential calculi by suitable functions.
The commutation relations for the differentials then no longer take the 
simple form (\ref{2-form-rels}). If two differential calculi are related by 
a (suitable) coordinate transformation,  they should be identified. A 
complete classification of two-dimensional differential calculi has not 
yet been achieved (see \cite{BDMH95} for partial results).  As a 
consequence of our definitions, the action of the $\ast$-operator can
be calculated on any basis of $\Omega^1({\cal A})$ if we know its
action on one basis.

\section{Generalized $\sigma$-models and conservation laws}
In case of the ordinary differential calculus on $\Rl^2$, the following
construction of conserved currents is due to Brezin et al \cite{Bre79}. 
In the form presented below, it also works for the noncommutative 
geometries introduced in the previous section. 
$\Gamma$ denotes an algebra of finite matrices with entries in $\cal A$ 
and $\Gamma^\ast$ the group of invertible elements of $\Gamma$. 
For $g \in \Gamma$ and
\be           \label{pure-gauge}
          A := g^{-1} \, \d g  
\ee
we consider the field equations 
\be 
       \d \ast A = 0         \label{sigma-feqs}
\ee
and refer to such a classical field theory as a {\em generalized
$\sigma$-model}. Since $A$ is a `pure gauge' we have 
\be                 \label{F}
       F := \d A + A A = 0  \; . 
\ee
Let $\Psi \in \Gamma$  and $D : \Gamma \to \O^1 \otimes_{\A} 
\Gamma$ the `exterior covariant derivative' given by
\be                \label{cov-deriv}
       D \Psi = \d \Psi + A \, \Psi  \; .
\ee
Using (\ref{ast1}), (\ref{sigma-feqs}) and (\ref{ast2}), we find
\be
       \d \ast (A^i{}_j \, \Psi^j{}_k) = \d (\Psi^j{}_k \ast A^i{}_j) 
    = (\d \Psi^j{}_k) \ast A^i{}_j = A^i{}_j \ast \d \Psi^j{}_k
\ee
and thus
\be               \label{dD}
      \d \ast D \Psi = D \ast \d \Psi \; . 
\ee
If there is one conserved current for a generalized $\sigma$-model,  
then an infinite sequence of conserved currents is obtained  as follows. 
Suppose $J^{(m)} \in \O^1 \otimes_{\cal A} \Gamma$ is conserved, i.e.,
\be  
       \d \ast J^{(m)} = 0   \; . 
\ee
If the first cohomology group of $\Omega({\cal A})$ is trivial and
provided that (\ref{astast}) holds, 
there exists $\chi^{(m)} \in \Gamma$ such that
\be 
         J^{(m)} = \ast \, \d \chi^{(m)} \; .         \label{dc}
\ee
Then 
\be 
         J^{(m+1)} := D \chi^{(m)}             \label{Dc}
\ee
is also conserved since
\be  
    \d \ast J^{(m+1)} = \d \ast D \chi^{(m)}
    =  D \ast \d \chi^{(m)} 
    =  D J^{(m)} = DD \chi^{(m-1)} = F \, \chi^{(m-1)} = 0   \; .
\ee
Starting with $\chi^{(0)} = I$, the unit matrix, this procedure indeed 
generates an infinite number of conserved currents. Let us introduce 
\be  
            \chi := \sum_{m=0}^\infty \kappa^m \, \chi^{(m)}
\ee
where $\kappa$ is a parameter. From (\ref{dc}) and (\ref{Dc}) we 
obtain
\be 
            \ast \, \d \chi^{(m+1)} = D \chi^{(m)}  \; . 
\ee
Multiplying by $\kappa^{m+1}$ and summing over $m$ leads to 
\be                         \label{lin-system}
            \ast \, \d \chi =\kappa \, D \chi   \; .               
\ee
The field equations (\ref{sigma-feqs}) are integrability conditions 
of the linear system (\ref{lin-system}). In a slightly more general
setting this will be shown in the following section.

\section{Another look at the linear system}
Let $A \in \Omega^1({\cal A}) \otimes_{\cal A} \Gamma$. Here $A$ is 
not assumed to have the form (\ref{pure-gauge}). We still use the 
definitions (\ref{F}) and (\ref{cov-deriv}), however. Let us consider
 a linear system of the form (\ref{lin-system}), i.e.,  $ \ast \, \d \chi 
=\kappa \, D \chi $.               
It implies
\be
  0 &=& \d ( \ast \, D \chi)^i{}_j  = \d \, \ast \, \d \chi^i{}_j 
        + \d ( \chi^k{}_j \ast A^i{}_k )    \nonumber \\
    &=& d \, \ast \, \d \chi^i{}_j +  A^i{}_k \ast \, \d \chi^k{}_j 
        + \chi^k{}_j \, \d \ast A^i{}_k  \nonumber \\
    &=& (D \ast \, \d \chi)^i{}_j + \chi^k{}_j \, \d \ast A^i{}_k  \; .
\ee
On the other hand,  (\ref{lin-system}) also leads to
\be 
        D \ast \d \chi = \kappa \, D^2 \chi = \kappa \, F \, \chi \; .
\ee
Hence
\be
  \chi^k{}_j \; \d \ast A^i{}_k = - \kappa \, F^i{}_k \, \chi^k{}_j \; .
\ee
We can now achieve $F = 0$ with the ansatz (\ref{pure-gauge}), i.e., 
$A = g^{-1} \d g$, as we did in the previous section. Then 
(\ref{sigma-feqs}) is the integrability condition of (\ref{lin-system}) 
which then depends on $g$. Alternatively, we can satisfy 
$\d \, \ast A = 0$ by setting $A = \ast \, \d g'$. Then $F = 0$ is
the integrability condition for the above linear system which now 
depends on $g'$. We should stress that in the two cases we are dealing 
with different linear systems and one should not expect the equations 
resulting from the two integrability conditions to be equivalent. In the 
following two examples, this turns out to be the case, however.
\vskip.3cm
\noindent
{\em Example 1.} Let us consider the differential calculus of example 2 
in section 2 with $b=0$ (so that elements of $\cal A$ should be 
$C^\infty$-functions of $t$) and $\ast \, \d t = \alpha \, \d x \, , \; 
\ast \, \d x = \beta \, \d t$ where $\alpha, \beta$ are constants 
different from zero. For $v \in {\cal A}$ we write $v_n(t) = v(n a, t)$ 
where $n \in \Ir$. Then
\be
  \d v_n = \d t \, \dot{v}_n + \d x \, {1 \over a} \, (v_n - v_{n-1}) 
           \; .
\ee 
The 1-form
\be
      A(n a, t) := \ast \, \d v_n = \alpha \, \dot{v}_n \, \d x
                       + {\beta \over a} \, (v_n - v_{n-1}) \, \d t   
\ee
has the `curvature' 
\be
   F(n a, t) = \lbrack \alpha \, \ddot{v}_n - {\beta \over a^2} \,  
               (1 + \alpha \, a \, \dot{v}_n) \, (v_{n+1} - 2 v_n 
               + v_{n-1}) \rbrack \, \d t \, \d x  \; .
\ee
The zero curvature condition $F = 0$ is then equivalent to
\be
  \lbrack \mbox{ln}( 1 + \alpha \, a \, \dot{v}_n) \rbrack \, \dot{ }
   =  {\beta \over a} \,   (v_{n+1} - 2 v_n + v_{n-1})  \; .
\ee
This equation is `dual', in the sense of an exchange of the roles of 
particles and interactions, and mathematically equivalent to that of the 
nonlinear Toda lattice equation, see \cite{Toda}, p. 18. The latter is
\be
  \ddot{u}_n = {\beta \over \alpha \, a^2} \, \left ( e^{u_{n-1} - u_n}
               - e^{u_n - u_{n+1}}  \right)   
\ee 
which is recovered from $\d \ast A = 0$ where now 
$A = e^u \, \d e^{-u}$, i.e., (\ref{pure-gauge}) with $g = e^{-u}$. See 
also \cite{DMH95}.                                                     
\vskip.3cm
\noindent
{\em Example 2.} We choose the differential calculus of example 3 
in section 2. For the 1-form
\be
   A := \ast  \, \d v = \lbrack \gamma \, ( v_t - {\eta \over 2} \, 
        v_{xx}) + \beta \, v_x \rbrack \, \d t - \gamma \, v_x \, 
        \d x \, ,
\ee
where $v \in {\cal A}$, the zero curvature condition $F = 0$ is
\be 
    w_t + {1 \over 2 \gamma} \, (\beta w)_x - {\eta \over 4} \, 
    \gamma \, (w^2)_x = 0
\ee
where $w = v_x$. On the other hand, from $\d \ast A = 0$, where now 
$A := e^u \, \d e^{-u}$, we obtain the same equation by setting
$w := \gamma \, u_x$.    
\vskip.3cm
\noindent
{\em Example 3.} Let us consider the $\nu = 0$ calculus of example 4 
in section 2. With $A = \ast \, \d v$ where $v \in {\cal A}$, the zero 
curvature condition is equivalent to\footnote{The function $\beta(t)$
can be absorbed by choosing a suitable `time' coordinate.} 
\be
   \bar{\partial} \dot{v}_n = - {\beta(t) \over 2 \gamma} \, \Delta v_n
   + {\gamma \over 2} \, \lbrack \dot{v}_{n-1} \, \partial_{+x} v_n
   + \dot{v}_{n+1} \, \partial_{+x} v_{n-1} \rbrack
\ee
where $v_n := v(n \mu, t)$ and
\be
   \bar{\partial} v_n := {1 \over 2 \mu} \, ( v_{n+1} - v_{n-1} ) 
   \, , \quad
   \Delta v_n := {1 \over \mu^2} \, ( v_{n+1} - 2 \, v_n + v_{n-1} )
   \; .
\ee
On the other hand, with $A = e^u \, \d e^{-u}$ the equation $\d \ast A 
= 0$ leads to
\be
     \dot{u}_{n+1} \, e^{u_n - u_{n+1}} - \dot{u}_{n-1} \, e^{u_{n-1} 
     - u_n}  =  {\beta(t) \over \gamma \, \mu} \, \lbrack
     e^{u_n - u_{n+1}} - e^{u_{n-1} - u_n} \rbrack  \; .
\ee
\vskip.3cm

These are just a few examples of integrable equations. The relevance
of the last two is unclear. They are included here mainly to illustrate 
the general method. So far we have restricted our examples to $g \in 
{\cal A}$ for simplicity. Generalizations to models where $g$ takes
values in some matrix group are easily obtained, as in the next example. 
 
\vskip.3cm
\noindent
{\em Example 4.} We generalize our example 3 in the sense just
mentioned. With $A = g^{-1} \d g$ (where $g \in \Gamma^\ast$) the
equation $\d \, \ast A = 0$ is equivalent to 
\be                    \label{int_eq_ex4}
   g_n^{-1} \, \dot{g}_{n+1} + (g^{-1}_{n-1}) \, \dot{ } \, g_n
   = - {\beta \over \mu \, \gamma} \, ( g^{-1}_n \, g_{n+1} - 
     g^{-1}_{n-1} \, g_n )  \; .
\ee
The linear system (\ref{lin-system}) can be expressed as follows
(when $\kappa \neq 0$),
\be
   ( g_{n+1} \, \chi_{n+1} ) \, \dot{ } &=&
   {1 \over \kappa} \, g_n \, ( \gamma \, \dot{\chi}_{n-1} 
   + \beta \, \partial_{+x} \chi_{n-1} )  \, ,        \\
   g_{n+1} \, \chi_{n+1} &=& g_n \, \lbrack (1- \gamma/\kappa) \,
   \chi_n + (\gamma/\kappa) \, \chi_{n-1} \rbrack  \; .
\ee
Introducing $\xi_n := (\kappa \, g_n \, \chi_n , \chi_{n-1})^T$ and
\be 
   L_n &:=& {1 \over \kappa} \, \left ( \begin{array}{cc}
          \kappa - \gamma & \gamma \, \kappa \, g_n \\
          g^{-1}_n        &  0  \end{array} \right )  \, \\
   M_n &:=& {1 \over \kappa - \gamma} \, \left ( \begin{array}{cc}
          \beta/\mu & - \kappa \, \lbrack \gamma \, \dot{g}_n 
           + (\beta/ \mu) \, g_n \rbrack \\
          - \lbrack g^{-1}_{n-1} \, \dot{g}_{n-1} 
          + \beta/ (\mu \, \gamma)
          \rbrack \, g^{-1}_{n-1} & \kappa \, \beta / (\mu \, \gamma)
          \end{array} \right ) 
\ee
(assuming $\kappa \neq \gamma$), the above system of equations can be
written as follows,
\be
   \xi_{n+1} = L_n \, \xi_n \, , \qquad 
   \dot{\xi}_n = M_n \, \xi_n  \; .
\ee
The integrability conditions, which are the $\sigma$-model field
equations, now take the form $\dot{L}_n + L_n \, M_n - M_{n+1}
\, L_n = 0$. We have derived a formulation of the complete integrability
of (\ref{int_eq_ex4}) in terms of a Lax pair. 
\vskip.3cm

In the way described in this section, and furthermore by choosing 
different differential calculi, we get a plethora of models which are 
integrable in the sense of section 3. These models need to be further 
investigated (in particular with respect to soliton solutions) and 
somehow classified.

\section{Conclusions}
We have introduced a generalization of $\sigma$-models in the
framework of noncommutative geometry.
Obviously our constructive method leads to a large set of new
completely integrable models. An interesting question is which of the 
known integrable models which are of interest in physics fit into this 
framework. For example, it has been shown in \cite{DMH95} (see also 
example 1 in section 4) that the nonlinear Toda lattice is a generalized
$\sigma$-model in the sense of section 3. Via the linear system 
(\ref{lin-system}) there is an integrable zero curvature model
associated with each generalized $\sigma$-model. This `duality'
turned out to coincide with a physical duality in case of the 
nonlinear Toda lattice.
\vskip.2cm

Our definition of generalized $\sigma$-models (and their duals) also
makes sense in more than two dimensions and the construction of
conserved currents in section 3 still works. 
The problem, however, is to find a $\ast$-operator satisfying 
(\ref{ast1}), (\ref{ast2}) and (\ref{astast}).   
It should also be noticed that, in more than two dimensions, our 
$\ast$-operator (which acts in the space of 1-forms) is no longer an 
analogue of the Hodge operator of Riemannian geometry. 
\vskip.3cm
\noindent
{\em Example.} Let us consider the ordinary differential calculus on 
$\Rl^n$. A $\ast$-operator is then determined by
\be
         \ast \, \d x^i = a^i{}_j \, \d x^j      
\ee
(using the summation convention). The symmetry condition 
(\ref{ast2}) takes the form
\be
        \omega'_k \, a^k{}_{\lbrack i} \, \omega_{j \rbrack}
     = \omega_k \, a^k{}_{\lbrack i} \, \omega'_{j \rbrack}
\ee
where $\omega = \d x^i \, \omega_i$ and the square brackets indicate
antisymmetrization of indices. In more than two dimensions ($n>2$), this 
condition is only satisfied for all 1-forms $\omega, \omega'$ if all the 
functions $a^i{}_j$ vanish.\footnote{Choose any three of the $n$ 
indices, like $\lbrace 1,2,3 \rbrace$, and evaluate the symmetry 
condition for $\omega_i, \omega'_j \in \lbrace \delta^1_k, \delta^2_k, 
\delta^3_k \rbrace$. This leads to $a^k{}_i = 0$ for all $i$ and for 
$k=1,2,3$. But since $\lbrace 1,2,3 \rbrace$ could number any triple of 
coordinates, we have $a^i{}_j = 0$ where $i,j = 1, \ldots, n$.}
Hence, there is no (generalized) $\sigma$-model in this case.
\vskip.3cm

The last example leaves us with a rather pessimistic impression
concerning the possibili\-ties of higher-dimensional generalized
$\sigma$-models. However, the situation may be different in case of 
other (noncommutative) differential calculi. The corresponding 
possibilities have still to be explored.

\end{document}